\documentclass{cjpsuppl_modified}
\usepackage[dvips]{epsfig}
\usepackage{amssymb}
\begin{document}

\title{Prospects for Scalar Leptoquark Discovery at the LHC}
\authori{V.\ A.\ Mitsou\footnote{Corresponding author; e-mail:
Vasiliki.Mitsou@cern.ch}}
\addressi{Instituto de F\'{i}sica Corpuscular (IFIC), CSIC –- Universitat de
Val\`{e}ncia, Valencia, Spain}
\authorii{N.\ Ch.\ Benekos}
\addressii{Max-Planck-Institut f\"{u}r Physik, Munich, Germany}
\authoriii{I.\ Panagoulias, Th.\ D.\ Papadopoulou}   \addressiii{Faculty of
Applied Mathematics and Physics, \\
National Technical University of Athens, Athens, Greece}
\authoriv{}    \addressiv{}
\authorv{}     \addressv{}
\authorvi{}    \addressvi{}
\headtitle{Prospects for Scalar Leptoquark Discovery at the LHC}
\headauthor{V.\ A.\ Mitsou {\it et al.}} \lastevenhead{V.\ A.\ Mitsou {\it et
al.}: Prospects for Scalar Leptoquark Discovery at the LHC} 
\refnum{}
\daterec{12 November 2004;\\. }
\suppl{A}  \year{2004} \setcounter{page}{1} \firstpage{1}
\maketitle

\begin{abstract}
The discovery potential of the ATLAS detector for scalar leptoquark pair
production at the LHC is discussed in this paper. The study is performed using
a parameterized yet realistic simulation of the ATLAS detector response for the
signal and the background. The channel $\rm LQ\,LQ\rightarrow(\ell q)(\ell q)$,
where $\rm\ell=e,\mu$, is investigated for the first two generations and the
decay mode $\rm LQ\,LQ\rightarrow(\nu_{\tau}b)(\nu_{\tau}b)$ for the third one.
In both cases, a preliminary mass reach is found to be ${\rm\sim1.3~TeV}$ for
three years of LHC running at low luminosity.
\end{abstract}

\section{Introduction and phenomenology}

Among possible new particles in physics beyond the Standard Model, leptoquarks
(LQs) are an interesting category of exotic colour triplets with couplings to
quarks and leptons. They are a generic prediction of Grand Unified Theories
\cite{GUT}, of composite models \cite{composite}, of technicolor schemes
\cite{technicolor}, of superstring-inspired $E_6$ models \cite{E6}, and of
supersymmetry with $R$-parity violation \cite{Rparity}.

Leptoquarks are colour triplets which couple to quarks and leptons via a
Yukawa-type coupling, $\lambda$, conserving the baryon and lepton numbers. In
the model of Buchm\"{u}ller, R\"{u}ckl and Wyler (BRW) \cite{BRW}, leptoquarks
couple to a single generation of Standard Model (SM) fermions via chiral Yukawa
couplings which are invariant under the ${\rm SU}(3)_C\times{\rm
SU}(2)_L\times{\rm U}(1)_Y$ symmetry group. Inter-generational mixing is not
allowed since neither flavour-changing neutral currents nor lepton flavour
violation, induced by this mixing, have been observed so far. Moreover,
leptoquarks coupling to both left- and right-handed electrons would mediate
rare decays \cite{rare}, hence leptoquarks couplings are assumed to be chiral.

There exist 14 species\footnote{If the assumption of chiral couplings is
dropped, this number reduces to ten.} of leptoquarks, differing by their spin
(scalars or vectors), fermion number $F=3B+L$, isospin and chirality of the
coupling. They have fractional electric charge ($\pm5/3,\pm4/3,\pm2/3$ and
$\pm1/3$) and decay into a charged or neutral lepton and a quark, according to
the decay modes $\rm{LQ\rightarrow\ell q}$ and $\rm{LQ\rightarrow\nu q}$. In
the most commonly used notation, $F=0$ scalar ($S$) and vector ($V$) species
are denoted as $S_{1/2}^L$, $S_{1/2}^R$, $\tilde{S}_{1/2}$, $V_0^L$, $V_0^R$,
$\tilde{V}_0$ and $V_1$, while the $F=2$ species are $S_0^L$, $S_0^R$,
$\tilde{S}_0$, $S_1$, $V_{1/2}^L$, $V_{1/2}^R$ and $\tilde{V}_{1/2}$
\footnote{Superscripts denote the chirality and subscripts indicate the isospin
of the particles.}. Decay widths for scalar and vector LQs of mass $M_{\rm LQ}$
are given by $\lambda^2M_{\rm LQ}/16\pi$ and $\lambda^2M_{\rm LQ}/24\pi$,
respectively. The leptoquark branching fractions
$\rm\beta\equiv\mathcal{B}(LQ\rightarrow\ell q)$ are predicted by the BRW model
and take the values 1, 0.5 and 0.

\begin{figure}[ht]
\centering\epsfig{file=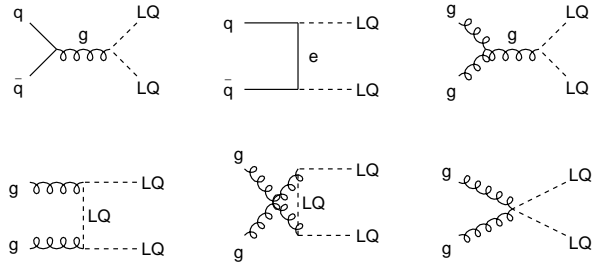,width=0.63\linewidth}
\vspace*{-0.4cm}
\caption{Leading-order diagrams of leptoquark pair production at LHC.}
\label{fig:feynman}
\vspace*{-0.3cm}
\end{figure}

In the following we consider pair production of scalar leptoquarks, which
proceeds through gg fusion and $\rm q\bar{q}$ annihilation via the diagrams
shown in Fig.~\ref{fig:feynman}. The production cross section is only slightly
dependent on $\lambda$, since it is dominated by gg fusion which does not
involve the $\rm\ell\!-\!q\!-\!LQ$ coupling. Therefore the cross section is not
strongly sensitive to the quark and lepton flavours to which the LQ couples,
implying that all species are produced at almost the same rate. More
information on LQ production at hadron colliders can be found in
Ref.~\cite{pheno}.

\section{Exclusion limits}

The current bounds on leptoquark production are set from Tevatron
\cite{Tevatron}, LEP \cite{LEP} and HERA \cite{HERA}. For first generation LQs,
the Tevatron experiments have set limits on scalars (coupling to eq) of $M_{\rm
LQ}>242~{\rm GeV}$ and corresponding vector LQ limits in the range from 233 to
345~GeV, depending on model assumptions. For LQs coupled to $\rm\nu q$ the
limit is set to 117~GeV. The LEP experiments have set bounds on $M_{\rm LQ}$
(approximately proportional to $\lambda$) which range from 165 to 917~GeV for
$\lambda=\sqrt{4\pi\alpha_{\rm em}}\simeq0.3$. The HERA experiments have set
lower mass limits in the range of $\sim250$ to 280~GeV for $\lambda=0.1$. In
addition, searches at the Tevatron and LEP have constrained leptoquarks coupled
to leptons and quarks of the second and third generations.

\section{Event and detector simulation}

In this study both the signal and the background events are generated with the
event generator PYTHIA~6.2 \cite{PYTHIA}, which provides a leading-order
calculation of the LQ production cross section. The CTEQ5M set is used for the
parton distribution functions and initial- and final-state radiation is
switched on. As far as the Yukawa coupling is concerned, we set
$\lambda=\sqrt{4\pi\alpha_{\rm em}}\simeq0.3$, leading to a total decay width
of $\Gamma\sim2~{\rm GeV}$ for $M_{\rm LQ}=1~{\rm TeV}$. This width is
overwhelmed by the calorimeter resolution precluding any measurement of it, yet
it allows one to treat the leptoquark as a resonance, i.e.\ it decays before
fragmentation and only the products of its decay are observed in the detector.
Hence the results obtained in this study are insensitive to $\lambda$ as long
as its value is consistent with resonant LQs, i.e.\ $\lambda\gtrsim10^{-6}$ for
$M_{\rm LQ}\gtrsim1~{\rm TeV}$.

The performance of the ATLAS detector \cite{ATLAS} was simulated with the fast
simulation package ATLFAST \cite{ATLFAST}. This simulation includes, in a
parameterized way, the main aspects related to the detector response: jet
reconstruction in the calorimeters, momentum/energy smearing for leptons and
photons, reconstruction of missing transverse energy and charged particles. It
is tuned to reproduce as well as possible the expected ATLAS performance
\cite{TDR}.

\section{Leptons plus jets channel ($\rm\ell\ell jj$)}

We have considered pair production of scalar leptoquarks of the first two
generations. Both LQs are assumed to decay to a charged lepton and a quark,
providing a topology with two high-$p_{\rm T}$ leptons and two high-$E_{\rm T}$
jets. Potential bounds in LQ mass are obtained for leptoquarks with branching
ratio $\beta=1$ or $\beta=0.5$.

\subsection{Background}

The background processes considered are characterized by the presence of two
isolated high-$p_{\rm T}$ leptons and at least two jets in the final state.
Before applying any cuts the main background source is the QCD processes.
Nevertheless this is eliminated completely after the requirement for two
isolated high-$p_{\rm T}$ leptons and in particular after selecting events with
high lepton-jet reconstructed invariant mass, $m_{\rm\ell j}$. The Drell-Yan
lepton pair production, $\rm
q\bar{q}\rightarrow\gamma^{\ast}/Z\rightarrow\ell^+\ell^-$, is another possible
background source but not as important for the LHC (pp collider) as it is for
Tevatron ($\rm p\bar{p}$ collider). It can be eliminated in a similar manner
after the high-$m_{\rm\ell j}$ cut is applied.

The cross sections of the surviving background processes are given in
Table~\ref{tb:bgd1}. The production of two leptons and two jets in the final
state has been forced in PYTHIA. For processes with large cross section, such
as Z+jet and $\rm t\bar{t}$, the generation production has been divided in
several $\hat{p}_{\rm T}$ intervals.

\begin{table}[h!t]
\caption{Cross section, expected number of events for $L=30~{\rm fb}^{-1}$
and number of generated events of background processes for the $\rm
LQ\,LQ\rightarrow\ell\ell jj$ channel.}\label{tb:bgd1}
\vspace{1mm}\small
\begin{center}
\begin{tabular}{|c|c|c|c|}
\hline
\raisebox{0mm}[4mm][2mm] {Processes} & $\sigma\times{\mathcal B}$ (pb) & \#
events & \# generated events \\ \hline\hline

\raisebox{0mm}[4mm] {Z+jets $\rm(\ell\ell jj)$, $\hat{p}_{\rm T}>20~{\rm GeV}$
} & 1380 & $4.14\cdot10^7$ & $2.8\cdot10^6$ \\ \hline

\raisebox{0mm}[4mm] {$\rm t\bar{t}$ $\rm(e\nu be\nu b)$} & 5.7 & $1.7\cdot10^5$
& $5\cdot10^5$ \\ \hline

\raisebox{0mm}[4mm] {$\rm t\bar{t}$ $\rm(\mu\nu b\mu\nu b)$} & 5.7 &
$1.7\cdot10^5$ & $5\cdot10^5$ \\ \hline

\raisebox{0mm}[4mm] {ZZ $\rm(\ell\ell jj)$} & 1.2 & $3.6\cdot10^4$ &
$2\cdot10^5$ \\ \hline

\raisebox{0mm}[4mm] {ZW $\rm(\ell\ell
jj)$} & 1.2 & $3.6\cdot10^4$ & $2\cdot10^5$ \\ \hline

\raisebox{0mm}[4mm][2mm] {WW $\rm(\ell\nu\ell\nu)$} & 3.3 & $10^5$ & $10^5$ \\
\hline

\end{tabular}
\vspace{-1mm}
\end{center}
\end{table}

\subsection{Analysis}

A first level of cuts is applied, requiring exactly two same-flavour,
opposite-sign leptons with $p_{\rm T}>30~{\rm GeV}$ and $|\eta|<2.5$, and at
least two jets with $E_{\rm T}>30~{\rm GeV}$ and $|\eta|<5$, for both signal
and background, in order to identify the kinematic variables sensitive to the
signal. Furthermore the selection criteria were optimized so that a
significance over $5\sigma$ is achieved for the maximum possible leptoquark
mass, retaining at the same time at least ten signal events for an integrated
luminosity of $\rm30~fb^{-1}$. The final cuts for both $\rm eejj$ and
$\rm\mu\mu jj$ channels are the following: two same-flavour, opposite-sign
leptons with $p_{\rm T}>100~{\rm GeV}$ and $|\eta|<2.5$; at least two jets with
$E_{\rm T}>70~{\rm GeV}$ and $|\eta|<5$; invariant mass of the two leptons
$m_{\ell\ell}>180~{\rm GeV}$, to exclude events from the $\rm
Z\rightarrow\ell\ell$ peak; $E{\rm _T^{miss}}<70~{\rm GeV}$, to suppress the
$\rm t\bar{t}$ background; sum of transverse energy deposited in the
calorimeters $\sum E{\rm _T^{calo}}>570~{\rm GeV}$; ratio $E{\rm
_T^{miss}}/\sum E{\rm _T^{calo}}<0.05$ and a mass window $|m_{\rm\ell j}-M_{\rm
LQ}|<100~{\rm GeV}$. In the $m_{\rm\ell j}$ distribution, only the two leading
jets and the two leptons are taken into account in the mass reconstruction.
Moreover, between the two possible combinations, the one providing the minimum
difference between the obtained values of $m_{\rm\ell j}$ is chosen.

\begin{figure}[ht]
\centering\epsfig{file=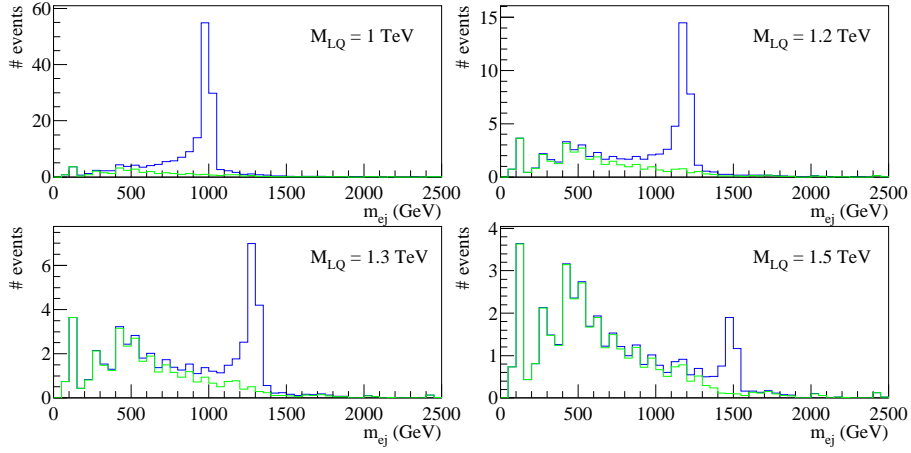,width=0.95\linewidth}
\vspace*{-0.3cm}
\caption{$m_{\rm ej}$ distributions for background (light green line)
and signal plus background (dark blue line), for various values of LQ mass and
for an integrated luminosity of $\rm 30~fb^{-1}$.}\label{fig:sig_bgd}
\end{figure}

The obtained $m_{\rm ej}$ distributions for the signal and the background,
after applying the final cuts, are shown in Fig.~\ref{fig:sig_bgd} for various
leptoquark masses and for the $\rm eejj$ channel. The cross section, the
corresponding numbers of events in the mass window, as well as the significance
achieved, are listed in Table~\ref{tb:signif1} for a branching ratio of
$\beta=1$. The distributions are normalized to an integrated luminosity of
$\rm30~fb^{-1}$, expected to be collected after three years of LHC running at
low luminosity.

Signal can be clearly observed over the background for up to $M_{\rm
LQ}\simeq1.3~{\rm TeV}$. For larger LQ masses, although the analysis is
practically background-free, the low statistics preclude any potential
observation of signal. Production cross section for the second generation
($\rm\mu\mu jj$ channel), is about 2\% lower, however, similar results are
obtained.

The leptoquarks studied so far decay with $\beta=1$ and correspond to the
following particles: ${\rm S}_0^R(-1/3)$, $\tilde{\rm S}_0(-4/3)$, ${\rm
S}_{1/2}^L(-5/3)$, ${\rm S}_{1/2}^R(-5/3)$, ${\rm S}_{1/2}^R(-2/3)$,
$\tilde{\rm S}_{1/2}(-2/3)$, ${\rm S}_1(-4/3)$ \footnote{The numbers in
parenthesis represent the electric charge of the particles.}. In the case of
leptoquarks with $\beta=0.5$, i.e.\ ${\rm S}_0^L(-1/3)$, ${\rm S}_1(-1/3)$, the
previous limit becomes less stringent; leptoquarks are observable for $M_{\rm
LQ}\lesssim1~{\rm TeV}$ with a significance of $S/\sqrt{B}\simeq15$ at $M_{\rm
LQ}=1~{\rm TeV}$. However this bound can be enhanced by searching for LQs in
the $\rm LQ\,LQ\rightarrow\ell\nu qq'$ channel.

\vspace*{2mm}

\begin{table}[h!t]
\vspace*{-4mm}
\caption{Signal ($\rm1^{st}$ generation) cross section, number of signal
and background events and significance for the $\rm eejj$ channel, for various
values of LQ mass and for $L=30~{\rm fb}^{-1}$.}\label{tb:signif1}
\vspace{1mm}\small
\begin{center}
\begin{tabular}{|c|c|c|c|c|}
\hline
\raisebox{0mm}[4mm][2mm] {$M_{\rm LQ}$ (TeV)} & $\sigma$ (fb) & Signal &
Background & $S/\sqrt{B}$ \\ \hline\hline

\raisebox{0mm}[4mm] {1.0} & 4.96 & 98.5 & 2.84 & 58 \\ \hline

\raisebox{0mm}[4mm] {1.2} & 1.33 & 22.0 & 2.43 & 14 \\ \hline

\raisebox{0mm}[4mm] {1.3} & 0.713 & 12.8 & 1.44 & 11 \\ \hline

\raisebox{0mm}[4mm] {1.5} & 0.223 & 3.62 & 0.376 & 5.9 \\ \hline

\end{tabular}
\vspace{-1mm}
\end{center}
\end{table}

\section{Missing energy plus jets channel ($\rm\nu\nu jj$)}

In the $\rm LQ\,LQ\rightarrow\nu\nu qq$ channel, both LQs are assumed to decay
to a neutrino and a quark, providing a topology with two high-$E_{\rm T}$ jets
and large missing transverse energy, $E{\rm _T^{miss}}$. For the first two
generations, the (huge) Z+jet $\rm(\nu\nu jj)$ background is irreducible making
thus the analysis unfeasible. In contrast, for a third-generation leptoquark
coupled to $\rm\nu_{\tau}b$, the b-tagging capabilities of the ATLAS detector
allow the signal to be disentangled from this background. Species with this
coupling are studied in the following. The branching ratio of the decay $\rm
\mathcal{B}(LQ\rightarrow\nu b)=1-\beta$ is taken to be equal to 1 or 0.5.

The standard ATLFAST b-tagging performance is assumed, i.e.\ a \mbox{b-tagging}
efficiency of 50\% with a fake rate of 9\% for c-quarks and 0.043\% for light
quarks. $p_{\rm T}$-dependent correction factors are also included.

\subsection{Background}

Besides the Z+jet $\rm(\nu\nu jj)$ production, main background sources are the
$\rm t\bar{t}$ $\rm(\ell\nu b\ell\nu b,$ $\rm\tau\nu b\tau\nu b)$, ZZ
$\rm(\nu\nu bb)$ and ZW $\rm(bb\ell\nu,\ bb\tau\nu)$ processes. All other
processes, listed in Table~\ref{tb:bgd2}, are eliminated by the b-tagging and
the lepton veto.

\begin{table}[h!t]
\caption{Cross section, expected number of events for $L=30~{\rm fb}^{-1}$
and number of generated events of background processes for the $\rm
LQ\,LQ\rightarrow\nu\nu bb$ channel.}\label{tb:bgd2} \vspace{1mm} \small
\begin{center}
\begin{tabular}{|c|c|c|c|}
\hline
\raisebox{0mm}[4mm][2mm] {Processes} & $\sigma\times{\mathcal B}$ (pb) & \#
events & \# generated events \\ \hline\hline

\raisebox{0mm}[4mm] {Z+jets $\rm(\nu\nu jj)$} & 22\,000 & $6.6\cdot10^8$ &
$7.5\cdot10^5$ \\ \hline

\raisebox{0mm}[4mm] {W+jets $\rm(\ell\nu bb,\ \tau\nu bb)$} & 28\,400 &
$8.5\cdot10^8$ & $1.5\cdot10^5$ \\ \hline

\raisebox{0mm}[4mm] {$\rm t\bar{t}$ $\rm(\ell\nu b\ell\nu b,\ \tau\nu b\tau\nu
b)$} & 51.6 & $1.5\cdot10^6$ & $5\cdot10^5$ \\ \hline

\raisebox{0mm}[4mm] {ZZ $\rm(\nu\nu bb)$} & 0.6 & $1.8\cdot10^4$ & $10^5$
\\ \hline

\raisebox{0mm}[4mm] {ZW $\rm(bb\ell\nu,\ bb\tau\nu)$} & 1.3 & $4\cdot10^4$ &
$10^5$ \\ \hline

\raisebox{0mm}[4mm] {ZW $\rm(\nu\nu jj)$} & 3.6 & $10^5$ & $10^5$
\\ \hline

\raisebox{0mm}[4mm][2mm] {WW $\rm(\ell\nu jj,\ \tau\nu jj)$} & 30.5 &
$9\cdot10^5$ & $10^5$ \\ \hline

\end{tabular}
\vspace{-1mm}
\end{center}
\end{table}

\subsection{Analysis}

Since in this channel leptoquarks decay to a $\tau$-neutrino and a b-quark,
producing two undetectable particles, the leptoquark mass is not reconstructed,
and therefore only an excess of events can be observed. The different topology
between signal and background events is taken into account by means of jet
separation angles. Nevertheless, the signal selection can be improved if event
shape variables, such as sphericity and aplanarity, are employed.

The final cuts applied to the event samples are the following: at least two
jets tagged as originating from b-quarks with $E_{\rm T}>70~{\rm GeV}$ and
$|\eta|<5$; missing transverse energy $E_{\rm T}^{\rm miss}>400~{\rm GeV}$;
veto on isolated leptons for suppression of W+jets and $\rm t\bar{t}$
background; and $m_{\rm jj}>180~{\rm GeV}$. In addition, since the two
leptoquarks are produced back-to-back, the angular distribution of the final
states should be constrained. Hence the azimuthal angle between the two leading
jets is required to be $30^{\circ}<\Delta\phi_{\rm j-j}<150^{\circ}$, and the
azimuthal angle between each of the two leading jets and the missing transverse
momentum is selected to be $\Delta\phi_{{\rm j}-p{\rm _T^{miss}}}>60^{\circ}$.

As inferred from Table~\ref{tb:signif2}, for $\beta=0$ ($\tilde{\rm
S}_{1/2}(1/3)$) leptoquark observation is feasible for LQ masses of up to
$\sim\!1.3~{\rm Tev}$. The corresponding exclusion limit at 95\% confidence
level that this analysis can set is $M_{\rm LQ}\lesssim1.5~{\rm TeV}$.
Moreover, if $\beta=0.5$ (${\rm S}_0^L(-1/3)$, ${\rm S}_1^L(-1/3)$) ---in which
case LQs also couple to a $\tau$-lepton and a top quark--- the ATLAS reach
using this channel reduces to $M_{\rm LQ}\simeq1~{\rm TeV}$ with a significance
of $\sim\!10$ for $M_{\rm LQ}=1~{\rm TeV}$.

\begin{table}[h!t]
\caption{Signal ($\rm3^{rd}$ generation) cross section, number of signal
and background events and significance for the $\rm\nu\nu bb$ channel, for
various values of LQ mass and for $L=30~{\rm fb}^{-1}$.}\label{tb:signif2}
\vspace{1mm}
\small
\begin{center}
\begin{tabular}{|c|c|c|c|c|}
\hline
\raisebox{0mm}[4mm][2mm] {$M_{\rm LQ}$ (TeV)} & $\sigma$ (fb) & Signal &
Background & $S/\sqrt{B}$ \\ \hline\hline

\raisebox{0mm}[4mm] {1.0} & 4.84 & 70.7 & 3.4 & 38 \\ \hline

\raisebox{0mm}[4mm] {1.2} & 1.28 & 21.3 & 3.4 & 12 \\ \hline

\raisebox{0mm}[4mm] {1.3} & 0.68 & 12.1 & 3.4 & 6.5 \\ \hline

\raisebox{0mm}[4mm] {1.5} & 0.21 & 3.9 & 3.4 & 2.1 \\ \hline

\end{tabular}
\vspace{-1mm}
\end{center}
\end{table}

\section{Discussion and conclusions}

The existence of TeV-scale scalar leptoquarks can be probed with the ATLAS
experiment at the LHC for all fermion generations in pair production channels,
extending the accessible LQ mass range by up to an order of magnitude compared
to currently running collider experiments. In the $\rm
LQ\,\overline{LQ}\rightarrow\ell^+\ell^-q\bar{q}$ mode the observation of a
first- or second-generation leptoquark is feasible for up to $M_{\rm
LQ}\simeq1.3~{\rm TeV}$ ($M_{\rm LQ}\simeq1~{\rm TeV}$) assuming $\beta=1$
($\beta=0.5$) with an integrated luminosity of $30~{\rm fb}^{-1}$. In the $\rm
LQ\,\overline{LQ}\rightarrow\nu_{\tau}\bar{\nu}_{\tau} b\bar{b}$ channel, on
the other hand, third-generation leptoquarks coupling to a $\tau$-neutrino and
a b-quark (b-antiquark) with $\beta=0.5$ ($\beta=0$) will be equally probed for
LQ masses of up to $1~{\rm TeV}$ ($1.3~{\rm TeV}$).

It has to be emphasized that this analysis treats leptoquarks from a purely
phenomenological point of view, i.e.\ the method is independent of the
underlying theory. To this respect, it concerns any particle decaying to a
lepton and a quark, e.g.\ squarks in some $R$-parity violating supersymmetric
scenarios where lepton number is not conserved.

The results presented here are not meant to be considered as final, but rather
represent an approximate yet substantial idea of the ATLAS potential as far as
leptoquark discovery is concerned. Further studies with GEANT-based simulated
events will shed light on the actual profile of the signal events as seen from
the ATLAS detector, possibly allowing more elaborate selection criteria to be
imposed. Moreover, if next-to-leading order calculation \cite{NLO} was employed
for the LQ production cross section, the reach is expected to be enhanced.
Study of other decay modes such as the `mixed' $\rm
LQ\overline{LQ}\rightarrow\ell q\nu \bar{q}'$ channel will also allow to
improve the reach.


\section*{Acknowledgements}

{\small This work has been performed within the ATLAS Collaboration and we
thank collaboration members for helpful discussions, and in particular the
convenors of the ATLAS Exotics working group, G.~Azuelos, S.~Ferrag and
G.~Brooijmans, for their comments and suggestions. V.A.M.\ would like to thank
the organizers of the {\em Physics at LHC} conference for the stimulating
environment they offered during the meeting. The work of V.A.M.\ is supported
by the EU funding under the RTN contract: HPRN-CT-2002-00292, {\em Probe for
New Physics}.


\end{document}